# Probing physical origin of anisotropic thermal transport in black phosphorus nanoribbons

*Yunshan Zhao[1], Gang Zhang[2], Mui Hoon Nai[3], Guangqian Ding[4], Dengfeng Li[4], Yi Liu[1], Kedar Hippalgaonkar[5], Chwee Teck Lim[3,6], Dongzhi Chi[5], Baowen Li[7,*], Jing Wu[5,*], John T L Thong[1,*]*

[1]Department of Electrical and Computer Engineering, National University of Singapore, Singapore 117583, Republic of Singapore

[2]Institute of High Performance Computing, Singapore, Singapore 138632, Republic of Singapore

[3]Department of Biomedical Engineering, National University of Singapore, Singapore 117576, Republic of Singapore

[4]School of Science, Chongqing University of Posts and Telecommunications, Chongqing, 400065, China

[5]Institute of Materials Research and Engineering, Agency for Science, Technology and Research, Singapore, 138634, Republic of Singapore

[6]Mechanobiology Institute, National University of Singapore, Singapore 117411, Republic of Singapore

[7]Department of Mechanical Engineering, University of Colorado, Boulder 80309, USA

Abstract

Black phosphorus (BP) has emerged as a promising candidate for next generation electronics and optoelectronics among the 2D family materials due to its extraordinary electrical/optical/optoelectronic properties. Interestingly, BP shows strong anisotropic transport behaviour because of its puckered honeycomb structure. Previous studies have demonstrated the thermal transport anisotropy of BP and theoretically attribute this to the anisotropy in both phonon dispersion relation and phonon relaxation time. However, the exact origin of such strong anisotropy lacks clarity and has yet to be proven experimentally. In this work, we probe the thermal transport anisotropy of BP nanoribbons (NRs) by an electron beam technique. We provide direct evidence that the origin of this anisotropy is dominated by the anisotropic phonon group velocity for the first time, verified by Young's modulus measurements along different directions. It turns out that the ratio of thermal conductivity between zigzag (ZZ) and armchair (AC) ribbons is almost same as that of the corresponding Young modulus values. The results from first-principles calculation are consistent with this experimental observation, where anisotropic phonon group velocity between ZZ and AC is shown. Our results provide fundamental insight into the anisotropic thermal transport in low symmetric crystals.

Keywords: Black phosphorus, Young's modulus, thermal conductivity, electron beam technique, first-principles calculation, anisotropy

Black phosphorus (BP) has attracted considerable attention as a promising two-dimensional (2D) material due to its high hole mobility[1, 2] and tunable bandgap[3-5]. Similar to graphene and other 2D materials, atomic layers in black phosphorus are stacked together through van der Waals interactions, making the material suitable for exfoliation. Layer-dependent bandgap has also been reported previously[2, 3] in BP like in the transition metal dichalcogenides (TMDCs). Because of its special puckered honeycomb structure formed by covalently-bonded phosphorus atoms, BP shows interesting angle-dependent transport properties[1, 2, 6]. Thus far, while extensive studies on electrical and optoelectronic applications of BP have been carried out, thermal transport measurement of BP is rarely explored.

Theoretically, it has been calculated that phonon transport in BP is anisotropic, where the thermal conductivity along the zigzag (ZZ) direction is larger than that along the armchair (AC) direction[7, 8] and an anisotropy around 3 was estimated by first-principles calculation[9]. Nevertheless, results from experimental studies on the thermal transport anisotropy of BP diverge significantly. By means of micro-Raman spectroscopy, Luo et al.[10] measured the thermal conductivity of BP thin films to yield an anisotropy ratio around 2 between ZZ and AC directions and this value dropped to ~1.5 for thinner films. Similarly, a thermal conductivity anisotropy ratio of 1.85 was obtained using the thermal bridge method and the anisotropy here was attributed to primarily the direction-dependent phonon dispersion and partially to the phonon scattering relaxation time[11]. Only the thermal conductivity measurement on bulk BP by the time-domain thermoreflectance (TDTR) technique[6, 12, 13] gives an anisotropy ratio up to 3, which is similar to that theoretically predicted. The slight difference in thermal conductivity anisotropy ratio comparing thin-film and bulk BP is possibly due to surface contamination like adsorbates and oxidation[10, 12], which may also be the key reason for the thermal conductivity anisotropy to vanish at low temperature[6, 11]. Even though these recent experimental works have measured the thermal transport anisotropy along ZZ and AC directions of BP, the behind-principle is either simply assumed or by theoretical calculation and the experimental verification of the origin of this anisotropy is still missing.

Considering $v_s \sim \sqrt{\frac{E}{\rho}}$, where $v_s$ is the speed of sound, $E$ the Young's modulus and $\rho$ the mass density[14, 15], a measurement of Young's modulus would provide direct information about the origin of phonon transport anisotropy, since the phonon group velocity is nearly equal to the speed of sound in the low frequency limit. This means we can directly link the anisotropic thermal conductivity of BP to its orientation-dependent Young's modulus, providing insight into the origin of thermal conductivity anisotropy of BP, as the Young's modulus is relatively insensitive to the thickness of BP[16]. Even though there are a few experimental works that aimed to measure the Young's modulus of BP[17-19], up to the present time the measured Young's modulus has not been used to explain the phonon transport anisotropy to elucidate the origin of thermal conductivity anisotropy from an experimental point of view.

In this work, we follow the BP NR fabrication process reported by Lee *et al*. [11] and study the anisotropic thermal conductivity of BP NRs with different thicknesses by an electron beam technique[15]. The anisotropic thermal transport of BP NRs is well accounted for by the Young's modulus measured along different directions using a three-point bending method[15]. The anisotropy ratio of thermal conductivity between ZZ and AC directions is similar to that of the Young's modulus along these two directions, implying that the thermal transport anisotropy is mainly due to the anisotropic phonon dispersion, and barely dependent on the phonon scattering relation time, which is contrary to predictions previously made by others that phonon scattering time would play a significant role in the phonon transport anisotropy of BP[11, 13]. The experimental observation is further supported by our first-principles calculation.

BP flakes were exfoliated on $SiO_2$/Si substrates in an Argon atmosphere glove box to avoid oxidation. A layer of PMMA was spun on top of the BP flakes for further protection and also served as the electron beam-resist for the following electron beam lithography (EBL) patterning.  In order to identify the ZZ and AC crystal directions in the BP flakes, polarized Raman spectroscopy (using WITEC alpha 300 system) was employed. The incident laser beam was kept parallel to the polarization of the collecting analyser during measurement. By rotating the sample from 0 to 360º, angle-sensitive Raman signals were obtained for different orientations. The ZZ and AC directions were identified by the intensity ratio between the peaks of the $A_g^1$ and $A_g^2$ phonon modes[11, 20], and the polarized angle dependent intensity ratio is shown in Supporting Information Figure S1. Figure 1 (a) shows the Raman sensitive peaks for ZZ and AC directions. The orientations of BP flakes with various thicknesses were characterized in this way before NR fabrication.

The BP flakes with identified orientations were then patterned by electron beam lithography (EBL), followed by deep reactive ion etching (RIE). BP NRs with three different thicknesses were studied, which were 106±4, 170±4.5 and 220±3.5 nm. Figure 1 (b) shows the SEM image of the fabricated BP NRs with thickness of 106±4 nm while data for the other BP NRs are provided in Supporting Information Figure S2. At least five NRs were fabricated for each direction, with the length kept at around 20 μm and width around 1 μm. The orientations for the BP NRs are indicated in Figure 1 (b).

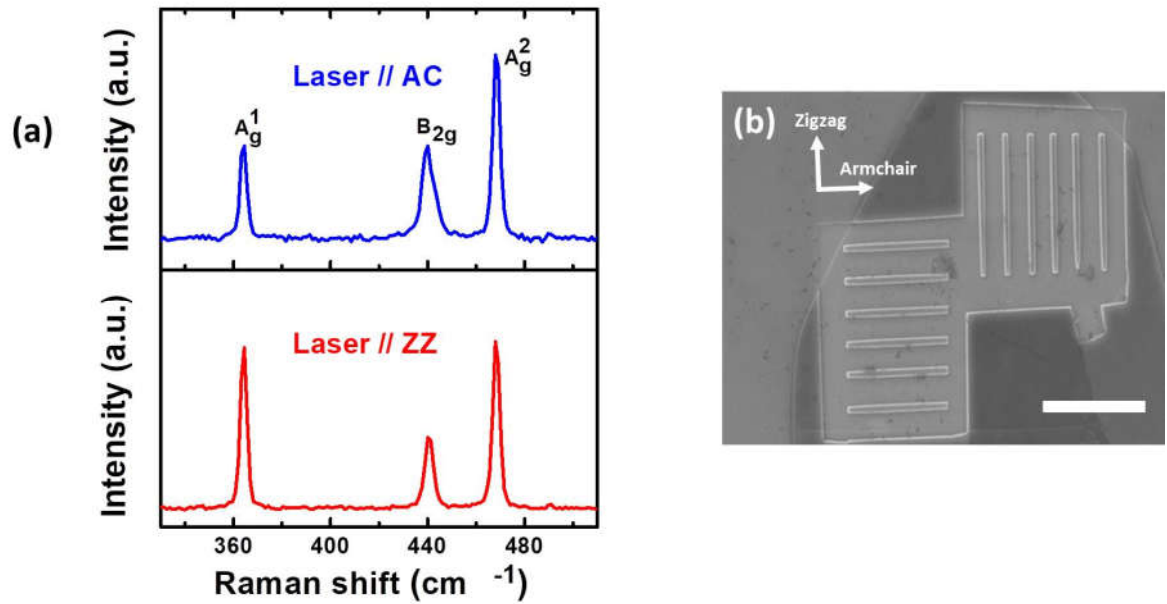

**Figure 1.** (a) Raman characterization of BP NRs along ZZ and AC directions. Raman sensitive peaks of $A_g^1$, $B_{2g}$ and $A_g^2$ are marked respectively. The top blue curve is for AC and bottom red curve for ZZ, respectively. (b) SEM image of fabricated BP NRs with ZZ and AC directions, indicated by polarized Raman spectroscopy. The length of BP NRs is around 20 μm and width is around 1 μm. The width is characterized by both SEM and AFM shown in Supporting Information Figure S2 and Figure S5. The scale bar is 15 μm.

Once the BP NRs were fabricated, the silicon substrate with BP NRs was transferred to the SEM chamber for thermal device fabrication, as well as fabrication of mechanical devices, which will be discussed in the next section. As shown in Figure 2 (a), a single BP NR was transferred onto a micro-electro-thermal systems (METS) device and the ends of the nanoribbons were deposited with platinum (Pt) to make good thermal contact[11, 15]. The transfer process is detailed in Supporting Information Figure S3. In total, six thermal devices were fabricated in this work, which are shown in Supporting Information Figure S4.

The thermal conductivity of BP NRs was measured by an electron beam technique, for which the measurement principle is discussed in our previous paper[15, 21]. Briefly, a focused electron beam is used as a heating source to heat up the NR and the temperature rises at the two ends of the NR, $T_L$ and $T_R$, were recorded by platinum resistance thermometers. By solving thermal transport formula at thermal steady state, the thermal conductivity is extracted as $\kappa = \frac{1}{A}\frac{1}{\frac{dR(x)}{dx}}$, where $R(x)$ is the thermal resistance distance $x$ from the starting point, and $A$ the cross-section area of BP NR. The cross-sectional area of the BP NR was determined by atomic force microscope (AFM) characterization discussed below. By scanning the electron beam along BP NR, the spatially-resolved thermal resistance can be obtained. Considering the BP NRs that were etched from the one same flake had the same cross-section area (same width and thickness), the slope of $R(x)$ would show the difference in the calculated thermal conductivity. As shown in Figure 2 (b), the slope of $R(x)$ for AC is much larger than that for ZZ, implying smaller thermal conductivity for BP along the AC direction.

The measured thermal conductivity of different thickness BP NRs is summarized in Figure 2 (b). With increasing thickness, the thermal conductivity of BP NRs increases due to decreased phonon-boundary scattering. From our study, the thermal transport anisotropy ratio for thickness of 106±4, 170±4.5 and 220±3.5 nm is around 2.24, 2.33 and 1.98, respectively, which is found to be similar to previous measurement results for BP of similar thickness[11]. Temperature dependent thermal conductivity of BP NRs with thickness of 106±4 nm was measured by the traditional thermal bridge method15, as shown in Figure 2 (d). With increasing temperature, the thermal conductivity of BP NR firstly increases and then decreases for both ZZ and AC directions. Anisotropic thermal conductivity between ZZ and AC directions persists at low temperature and the anisotropy ratio is invariant with temperature, as shown in Figure 2d, which disagrees with the observation in reference[11] that the thermal conductivity in both ZZ and AC directions merges at temperature close to 40K. For BP NR of the same thickness (106±4 nm), the thermal conductivity obtained by thermal bridge method is found to be a little smaller than that obtained by electron beam technique, as the thermal contact resistance is avoided in the electron beam technique measurement[15, 22].

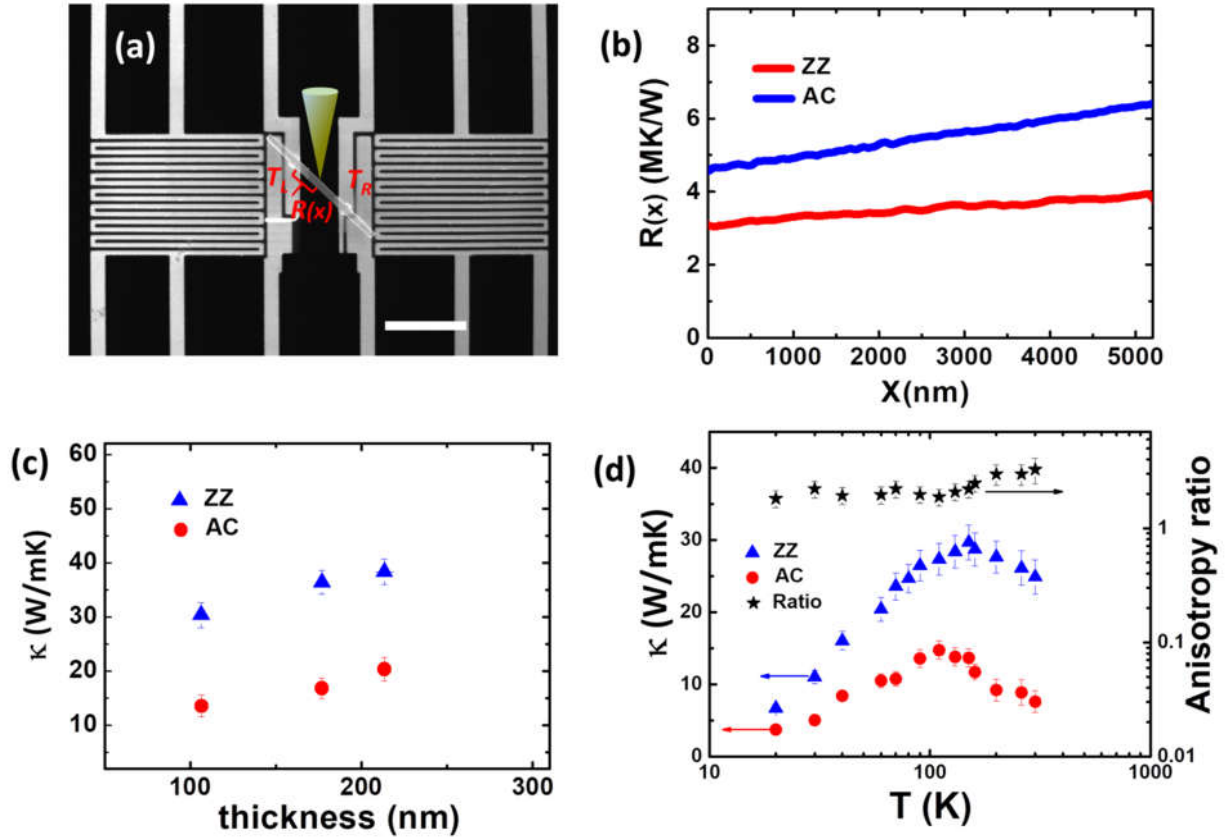

**Figure 2.** (a) SEM image consisting of two suspended PTR membranes and a bridging BP NR. $T_L$ and $T_R$ is temperature rise in the left and right PRT membrane, respectively. $R(x)$ is the thermal resistance distance $x$ from the starting point. The grey cone represents the electron beam. The scale bar for (a) is 10 μm. (b) The cumulative thermal resistance vs scanning distance during electron beam measurement. The red and blue curve is for AC and ZZ directional BP NRs with the same dimension. The thermal conductivity is extracted as $\kappa = \frac{1}{A}\frac{1}{\frac{dR(x)}{dx}}$, where $A$ is the cross section area of BP NRs. (c) Thermal conductivity of ZZ and AC directional BP NRs measured by electron beam technique. For each thickness, thermal conductivity of both ZZ and AC directions is measured. The red triangle and blue circle is for thermal conductivity of ZZ and AC, respectively. (d) Temperature dependent thermal conductivity of BP NRs with thickness of 106±4 nm, which is measured by thermal bridge method. The temperature-independent anisotropy ratio between ZZ and AC directions is shown as well.

BP NRs from the same batch of etched samples were used for fabricating devices to measure the mechanical properties. The detailed fabrication process is described in Supporting Information Figure S3. As illustrated in the sketch of Figure 3 (a), the Young's moduli of BP NRs were measured by the commonly-employed three-point bending method using a cantilever mounted on an AFM[15, 17, 23]. Holes with diameter of 7 μm and depth of 2 μm were etched in the silicon substrate using focused ion beam (FIB) milling. The BP NR was subsequently transferred onto one of the holes using a nano-manipulator tip in an SEM, followed by Pt deposition to fix the two ends.

The experiments were performed using a Dimension Icon AFM (Bruker). Force curves were obtained using the PeakForce Quantitative Nanomechanical Mapping (QNM) mode of the AFM. A cantilever (OTR8, Bruker) with a nominal spring constant of 0.57 N/m and tip radius of 15 nm was used. The maximum force was kept at 45 nN to ensure the NR undergoes elastic bending. A scan rate of 0.4 Hz was used, allowing 256 x 256 force curves to be captured over an area of 12x12 um2. A series of force curves across the midspan of the suspended NR was extracted using the analysis software and the largest measured deflection was used for the calculation of the Young's modulus. Figure 3 (c) shows these typical force vs displacement curves obtained at the mid-point of the AC and ZZ BP NRs. The simplified clamped-clamped beam equation which relates the loading force, F, to the beam deflection, $\delta$, was used as[17]

$$F = \frac{\pi^4 E w t^3}{6L^3}\delta + \frac{\pi^4 E w t}{8L^3}\delta^3$$

where $E$ is the Young's modulus, $w$ is the width, $t$ is the thickness and $L$ is the length of the suspended NR. The dimensions of the NR were determined from the topography profile obtained from AFM imaging and a representative cross-section profile of the BP NR is shown in Supporting Information Figure S5. The measured width of ~ 990±20 nm and height of ~106±4 nm was used for thermal conductivity calculation mentioned previously. The Young's moduli of the AC NR (43.6±3 GPa) and ZZ NR (89.4±3 GPa) are plotted with their thermal conductivity data in Figure 3 (d). It was observed that the thermal conductivity anisotropy ratio (~2.24) between ZZ and AC is similar to that of its Young's modulus data (~2.05).

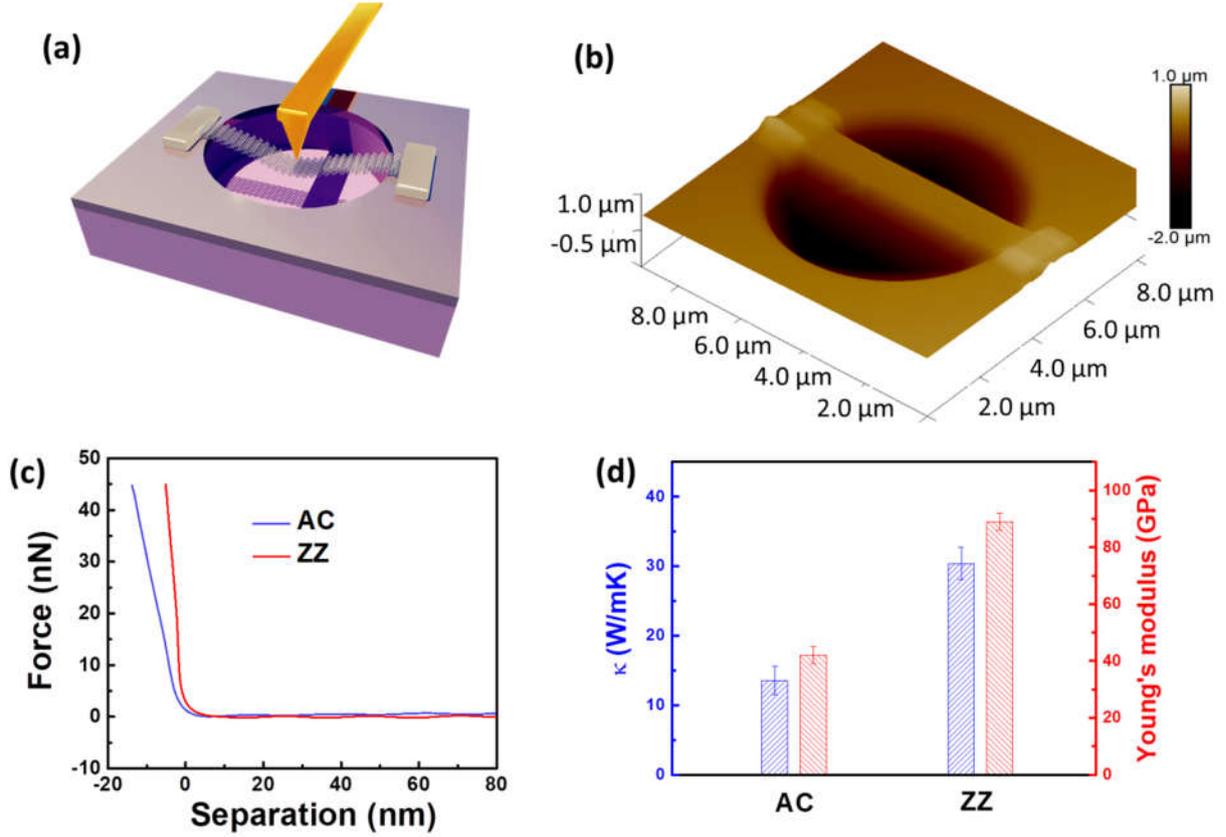

**Figure 3.** (a) Schematic diagram of the three-point bending measurement performed using an AFM cantilever. (b) 3D AFM image of a representative BP NR used for Young's modulus measurement. (c) Load vs displacement curves of the ZZ and AC NRs obtained at the midpoints. (d) Thermal conductivity and Young's modulus values of the BP NRs (106±4 nm thickness). The thermal conductivity anisoropy ratio (~2.24) between ZZ and AC is similar to that of Young's modulus (~2.05).

To better understand the relation between the measured Young's modulus and thermal conductivity, we recall the kinetic theory expression for thermal conductivity[15, 24], which is

$$\kappa = \frac{1}{3} C v l,$$

where $C$ is the volumetric specific heat, $v$ the average phonon group velocity and $l$ the phonon mean free path, which is equal to $v\tau$, where $\tau$ is an average phonon relaxation time. Considering the bulk-like volumetric specific heat for BP NRs, the thermal conductivity $\kappa$ is proportional to $v^2$, if the relaxation time is considered to be isotropic in both ZZ and AC directions[10]. On the other hand, the phonon group velocity is almost equal to the speed of sound, $v_s$, in the low frequency regime[15]. Since $v_s^2$ is proportional to $E$[14, 15] (Young's modulus), the Young's modulus difference between ZZ and AC directions could explain the difference observed in their thermal conductivity. Interestingly, this is confirmed by our experimental results that the thermal conductivity anisotropy ratio between ZZ and AC is similar to the ratio of the Young's modulus for the measured BP NRs. Our work provides the first experimental evidence for the origin of this thermal conductivity anisotropy of BP by measuring its interesting mechanical property.

According to previous theoretical work[25, 26], the Young's modulus anisotropy ratio of single layer BP between ZZ and AC directions is nearly constant in the temperature range of 10K-400K. If the phonon transport difference for ZZ and AC is indeed due to the direction dependent phonon group velocity, the thermal conductivity anisotropy ratio should be invariant with temperature, which is confirmed by the temperature-dependent thermal conductivity measurement of BP NR for both ZZ and AC directions in the previous section. The ratio of thermal conductivity along ZZ and AC directions at any temperature remains around 2. This anisotropy value is similar to that of the Young's modulus measured at room temperature. Therefore, the phonon transport anisotropy for BP mainly comes from the anisotropic phonon group velocity along ZZ and AC, verified by its Young's modulus measurement. If phonon relaxation time were to play a dominant role in the phonon transport anisotropy, the anisotropy ratio of thermal conductivity would be temperature dependent, which is not what we observed, implying minimal effect from the phonon relaxation time.

The reported thickness-dependent thermal conductivity results of BP of various dimensions are summarized in Figure 4 (a), which include those from BP thin films[10, 12, 27], nanosheets[28] and BP NRs[11], and the thickness of all these BP nanostructures is less than 1 μm. Even though the thermal conductivity is measured by different techniques, like time-domain thermoreflectance (TDTR)[12], thermal bridge[11, 27, 28], micro-Raman spectroscopy[10] and electron beam technique (this work), the overall trend is well captured by the size dependent thermal conductivity of phosphorene obtained from first-principles calculation[8]. Apparently, the size-dependent thermal conductivity of BP nanostructures implies that the phonon boundary scattering plays a role for a dimension smaller than 1 μm. For each thickness, the thermal conductivity anisotropy ratio between ZZ and AC is constant around 2. This ratio is comparable to that of Young's modulus between ZZ and AC summarized in Figure 4 (b), where the first-principles calculation[25, 29] and experimental measurement[17, 19, 30] results are compared. The overall simulated Young's modulus, even for single layer phosphorene, is comparable to that of experimentally measured data and its ratio between ZZ and AC is around 2.2±0.6. This anisotropy ratio is comparable to the speed of sound ratio that is derived in reference[6, 31], which is $v_{s,zz}^2/v_{s,AM}^2 \approx 2$. For both thermal conductivity and Young's modulus data measured in this work, they are consistent to work previously reported. According to reference[17], the Young's modulus of BP is nearly independent of its size due to week interlayer interactions, which could be observed as well in Figure 4 (b).

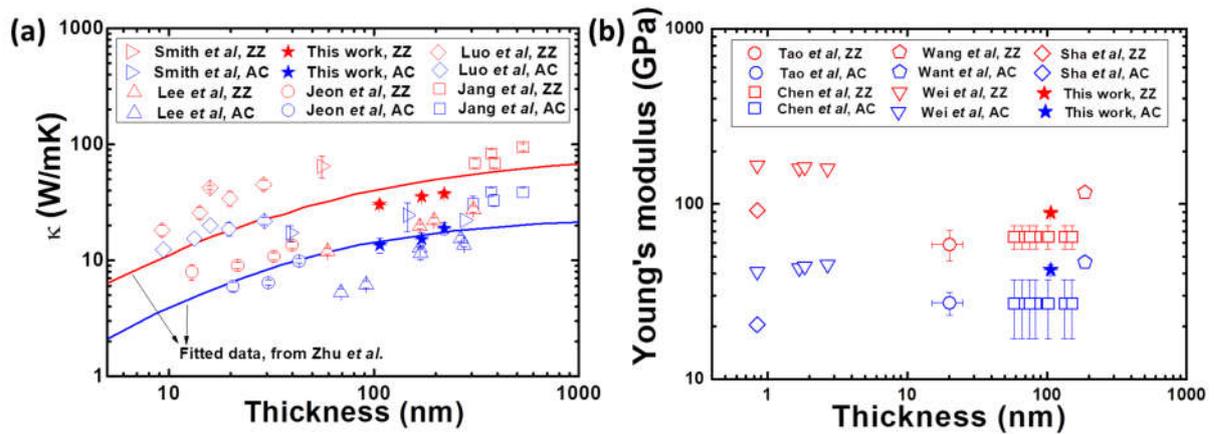

**Figure 4.** (a) Summary of thermal conductivity dependent on thickness for BP thin film[10, 12, 27], BP nanosheets[28] and BP NRs[11] (thickness is less than 1 μm). Thermal conductivity along ZZ and AC is in red and blue symbol, separately. Here the thermal conductivity measured by different techniques are compared, which are time-domain thermoreflectance (TDTR)[12], thermal bridge[11, 27, 28], micro-Raman spectroscopy[10] and electron beam heating technique (this work). First-principles calculations of size dependent thermal conductivity of phosphorene[8] are shown in red and blue curve for ZZ and AC. (2) Thickness dependent Young's modulus measured in this work and from literatures, which include the first-principles calculation[25, 29] and experimental measurement[17, 19, 30] results. Young's modulus along ZZ and AC is in red and blue symbol, separately.

To better understand the phonon transport anisotropy of BP NRs, we carried out first-principles calculation, where the phonon dispersion along both ZZ and AC was calculated and shown in Figure 5. We perform the structural optimization within the framework of density function theory (DFT) using projector-augmented-wave (PAW) pseudopotentials[32] and Perdew–Burke–Ernzerhof (PBE) exchange correlation functionals[33] as implemented in Vienna *ab initi*o simulation package[34]. For BP, the sound velocity is defined as the slope of longitudinal acoustic phonon branches at Gamma point[9], which is shown as dashed lines in Figure 5. Based on our calculation, the acoustic phonon modes dominate in phonon transport and it is found that $v_{zz}^2/v_{AC}^2$ is around 2.375, which is comparable to the thermal conductivity ratio along ZZ and AC directions measured in this work. It is thus eliminated that directional phonon relaxation time would play any role in the phonon transport anisotropy. It is noted as well that the value of $v_{ZZ}^2/v_{AC}^2$ is similar to the measured Young's modulus ratio between ZZ and AC directions. Therefore, our theoretical calculation supports that the phonon transport anisotropy in BP is primarily due to the anisotropic phonon dispersion with minimal effect from phonon relaxation time, as posited in previous works[6, 10].

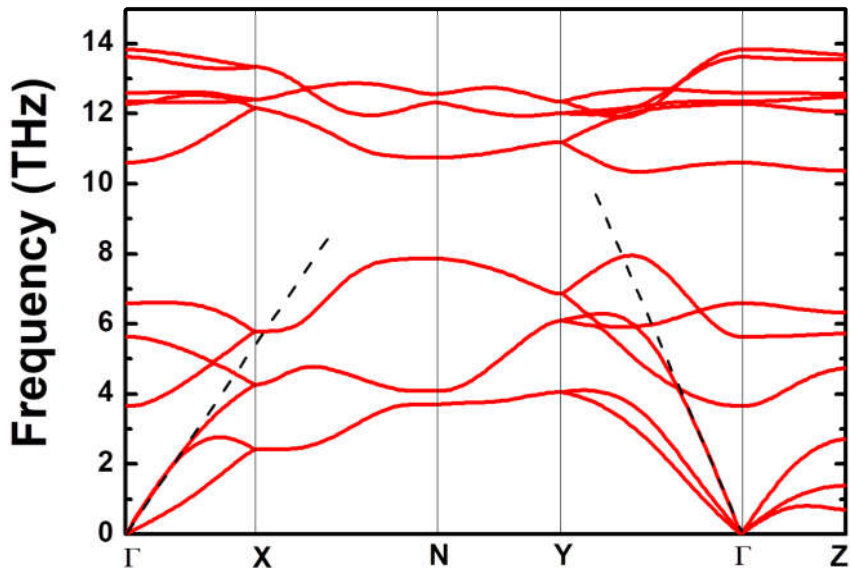

**Figure 5.** Calculated phonon dispersion of bulk black phosphorus. Γ-X corresponds to armchair direction of the primitive cell, and Γ-Y denotes the zigzag direction. According to paper[9], the sound velocity is equal to the slope of longitudinal acoustic phonon branches at Gamma point, which is shown as dashed straight lines.

To conclude, we experimentally reveal the origin of the anisotropic thermal transport in BP for the first time by linking its anisotropic phonon group velocity to the Young's modulus. The thermal conductivity of BP NRs of various thicknesses is measured by an electron beam technique and the thermal conductivity anisotropy ratio between ZZ and AC is comparable to that of Young's modulus data along these two directions, as measured by three-point bending method. Results from first-principles simulation show that the anisotropy is dominated by the anisotropic phonon dispersion relation along different directional angles with minimal effect from the phonon relation time. Our work provides a new direction to explore nanoscale phonon transport by studying the interesting macroscopic mechanical property, which aids to better thermal management at nanoscale as well.

## Supporting Information

Supporting Information is available from the Wiley Online Library or from the author.

## Acknowledgements
The authors gratefully acknowledge funding from A*STAR SERC, Grant No. 1527000015.